# Trion-based High-speed Electroluminescence from Semiconducting Carbon Nanotube Films

*Hidenori Takahashi[†,‡], Yuji Suzuki[†,‡], Norito Yoshida[†], Kenta Nakagawa[†,§] and Hideyuki Maki[†,∥,*]*

[†]Department of Applied Physics and Physico-Informatics, Keio University, Yokohama 223-8522, Japan.

[§]Kanagawa Institute of Industrial Science and Technology (KISTEC), Ebina 243-0435, Japan.

[∥]JST, PRESTO, Kawaguchi, 332-0012, Japan

ABSTRACT

High-speed light emitters integrated on silicon chips can enable novel architectures for silicon-based optoelectronics, such as on-chip optical interconnects and silicon photonics. However, conventional light sources based on compound semiconductors face major challenges for their integration with the silicon-based platforms because of the difficulty of their direct growth on a silicon substrate. Here, we report high-brightness, high-speed, ultra-small-size on-chip electroluminescence (EL) emitters based on semiconducting single-walled carbon nanotubes (SWNTs) thin films. The peaks of the EL emission spectra are 0.2-eV red-shifted from the peaks of the absorption and photoluminescence emission spectra, which suggests emission from trions. High-speed responses of ~ 100 ps were experimentally observed from the trion-based EL emitters, which indicates the possibility of several-GHz modulation. The pulsed light generation was also obtained by applying pulse voltage. These high-speed and ultra-small-size EL emitters can enable novel on-chip optoelectronic devices for highly integrated optoelectronics and silicon photonics.

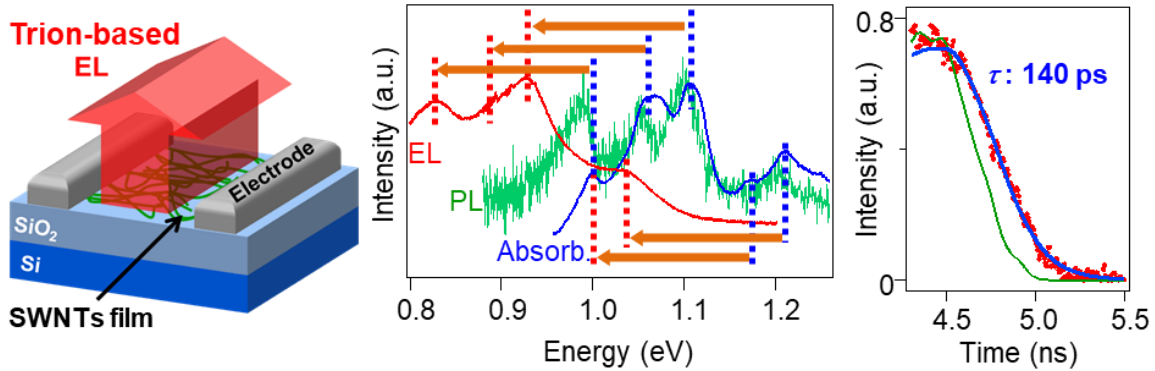

TEXT

**INTRODUCTION**

Light emitters on silicon chips are among the most important elementary devices for integrated optoelectronics and photonics, along with optical interconnects and silicon photonics. However, current light emitters based on compound semiconductors face challenges in their integration on silicon chips, because thin films of compound semiconductors are difficult to be directly formed on silicon substrates. Nanocarbon materials, such as carbon nanotubes (CNTs), are promising candidates for optoelectronic light sources on silicon chips, because they can be easily formed on silicon substrate. Two types of light emitters, blackbody emitters based on Joule heating[1–4] and electroluminescence (EL) emitters based on electron-hole recombination in semiconducting single-walled carbon nanotubes (SWNTs)[5–14], have been previously reported as nanocarbon-based emitters on silicon chips. Although these emitters demonstrated the steady-state light emission under DC bias voltage in most studies, high-speed light modulation is required for optical communications. Recently, high-speed modulation at ~ 1 GHz has been demonstrated for an individual-CNT[15] and CNT-film blackbody emitters.[2,3] On the other hand, for EL emitters, only slow modulation at ~ 100 Hz has been demonstrated[14], and there are no reports on high-speed EL emission.

EL emitters exhibit peak-shaped emission spectra determined by the bandgap of semiconducting SWNTs, which are different from the broad emission spectra of blackbody emitters. Light emission based on the electron-hole recombination is expected to exhibit higher quantum efficiency than that based on the blackbody radiation (e.g., $10^{-3}\%$[2] and $10^{-1}\%$[16]), as high quantum efficiency of ~ 20% has been demonstrated in photoluminescence (PL) measurements.[17] In addition, as chirality-enriched SWNTs are commercially available, semiconducting SWNTs can be easily formed on silicon wafers by a drop-casting or spin-coating process.[3,14,15,18–21] Furthermore, EL emitters are promising candidates for integrated light sources on silicon chips for optical communications with optical fibers and silicon photonic devices, because the EL spectra lie in the near-infrared (NIR) region, which includes telecommunication wavelengths.[15,22,23] However, the high brightness and high-speed EL emitters with stable performance have not been demonstrated to date.

In this paper, we report high-speed EL emitters using semiconducting SWNTs thin films in NIR region including telecommunication wavelength. By using spin-coated SWNTs films, we fabricated high-speed modulatable emitters with high emission intensity, high fabrication yield, and stable performance, compared with the EL emitters based on an individual SWNT. The electrodes of the EL emitters were designed using coplanar transmission lines to show characteristic impedance for the high-speed operation under high-frequency bias voltage.[2,24] The mechanisms of the EL emission in this emitter were elucidated by the emission spectrum and electric properties, which show the luminescence from trions generated by impact excitation. In trion emission under the current injection excitation, fast relaxation of ~ 100 ps or less was revealed by time-resolved measurements. We demonstrated 120-ps-width pulsed light emission under a pulsed voltage input, as well as the high-speed continuous modulation at 1 GHz.

**RESULTS AND DISCUSSION**

The schematic picture of the fabricated device is shown in Fig. 1a. In this study, we fabricated two types of devices: DC and high-speed devices. In DC devices, the gate voltage can be applied through p++ Si substrate, and the device can emit light under DC voltage. In high speed devices, the gate voltage cannot be applied, but the device can emit light under high frequency input because of the 50-Ω characteristic impedance of the device obtained by using undoped Si substrate. For DC devices, 30 × 30 µm square holes of the resist were patterned on the substrate, and the semiconducting SWNTs (SWeNT: SG76) purified by the CoMoCAT method were drop-casted and spin-coated; subsequently, square-shaped semiconducting SWNTs thin films were formed by the lift-off process (Fig. 1b). From the absorption spectrum, (7,6)-chirality SWNTs mainly constituted the films; other (8,6), (8,7), (7,5), and (10,2)-chirality SWNTs were also present.[25–28] The DC device is a typical two-terminal device, which can emit light by applying a bias voltage between the source and drain electrodes; a back-gate voltage can also be applied via the substrate. The source and drain electrodes with the gap of 500 nm and width of 30 µm were formed on the semiconducting SWNTs thin film (Fig. 1b). As the gap between the electrodes is shorter than the length of the used SWNT (~1 µm), the current flows via one or no more than several SWNTs. We note that our EL emitters are different from that of the previous study, in which the current flows via many SWNTs in SWNT networks between several-tens-µm electrode gap.[10,11,29–34]

For high-speed devices, a coplanar waveguide, which has a 50-Ω characteristic impedance, with a shape as shown in Fig. 1c, was formed on an undoped Si substrate, and high frequency voltage can be efficiently applied to the SWNTs thin film.[2,24] These high-speed devices were fabricated by the same fabrication processes as the DC device, except for the electrode shape and substrate type. A rectangular or pulse voltage was applied to the signal line to obtain the high-speed modulated EL emission in the high-speed modulation measurements.

The $I$-$V_{ds}$ characteristics of the fabricated DC device are shown in Fig. 1d. The DC bias voltage ($V_{ds}$) dependence of the current exhibits roughly ohmic behavior at a low bias voltage. The two-terminal resistance of this device, which was fabricated with purified semiconducting SWNTs, is 10–100 times higher than that of the device with unpurified SWNTs because of the high Schottky barrier at the contact and low carrier concentration due to little presence of metallic SWNTs in the semiconducting SWNT films. The $I$-$V_{ds}$ characteristics also show slight superlinear behavior, which is due to the current increase by the reduction of the Schottky barrier at high voltages. The gate voltage ($V_g$) dependence of the current at constant $V_{ds}$ (Fig. 1d inset) shows the p-type characteristic, in which the current increases in the negative $V_g$ region. Generally, in a field effect transistor with an individual semiconducting SWNT, the current is completely suppressed in the positive $V_g$ region; however, the current is not suppressed in our device with semiconducting SWNTs thin films. It can be explained by the following two mechanisms: one is that the electric field from the back gate is screened by a number of SWNTs, and the effect of the gate voltage is reduced compared with the device with an individual SWNT; the other is that the current leaks through minute quantity of metallic SWNTs contained in the semiconducting SWNTs thin films. Bright emission from the SWNTs between the electrodes was observed with a near-infrared (NIR) camera, as shown in Fig. 1e. In the camera images, spot-like light emission is observed because of the existence of particularly bright SWNTs. The reason is that the bright SWNTs show low contact resistances or small Schottky barriers because of the good contact between the SWNTs and electrodes.

Figure 2a shows the emission spectrum at $V_g = 0$ V under DC bias voltage $V_{ds}$. Several emission peaks are observed at the wavelength range from 1.1 to 1.6 µm, and the emission spectrum contain five peaks, which correspond to each of the band gaps of the SWNTs with different chirality in the purified semiconducting SWNTs. Figure 2b shows the absorption, PL, and EL spectra of the purified semiconducting SWNTs. The peaks of the absorption and PL emission, which correspond to the (8,7), (8,6), (7,6), (10,2), and (7,5) chiralities, are observed at almost similar wavelengths.[35–38] On the other hand, the peaks of the EL emission are 0.2 eV red-shifted from the peaks of the absorption and PL emission. As was previously reported, the EL emission energy from the exciton recombination coincides with the PL emission energy for the EL emitters based on the p-n junction; hence, the EL emission from our device is evidently different from the exciton states emission.[39,40] Such a red shift of the EL emission compared with the PL can be explained by the EL emission from the trion states.[20,37,41–47] As the emitter has a p-type characteristic, as shown in Fig. 1d, the emission is produced by the impact excitation mechanism. In the impact excitation mechanism, excitons with the energies lower than the band gap are formed continuously by the kinetic energy of the holes accelerated within the mean free path; hence, the local exciton density is increased. In addition, a large number of holes, which are required for trion formation, is efficiently injected because of the p-type characteristic of the device. Therefore, in the light emitter based on the impact excitation mechanism, trions are efficiently generated by forming both excitons and holes with high density. In previous studies[8,28,32,34-39], the energy state of the trion was reported 0.2 eV lower than the exciton states, which is consistent with our observation results shown in Fig. 2b suggesting trion formation. We note that the full width at half maximum (FWHM) of the emission peak slightly increases with increasing bias voltage $V_{ds}$ (Fig. 2a inset). It can be attributed to two factors: broadening of the trion binding energy due to the change of carrier interactions by increasing the number of excitons generated by the impact excitons at a high bias voltage[48–50] and increase of the carrier-carrier or carrier-phonon scattering by increasing the number of carriers or by increasing the temperature by Joule heating.[20,51]

Figure 2c shows the $V_{ds}$ dependences of the emission spectrum and total emission intensity. The emission intensity increases exponentially with increasing $V_{ds}$. When the thermal excitation is ignored, the excitation rate $P(V_{ds})$ by the impact exciton mechanism is given by the following equation:[3,10,13,52]

$$P(V_{ds}) = P_0 \exp\left(-\frac{E_{th}}{e\lambda_{OP}\alpha V_{ds}}\right) \qquad (1)$$

where $E_{th}$ is the excitation energy of the exciton ($E_{th}$ = 1.1 eV from the PL emission intensity), $\lambda_{OP}$ =14×$d$ is the optical phonon mean free path dependent on the diameter $d$ of the SWNT ($d$ = 0.83 nm in our device). Assuming that the emission intensity as a function of $V_{ds}$ is proportional to the excitation rate, we fitted the experimental $V_{ds}$ dependence of the emission intensity using $P_0$ and $\alpha$ as fitting parameters. As shown in Fig. 2c, the fitted curve is in good agreement with the experimental result, and $\alpha = 2.8 \times 10^{-3}$ obtained by fitting is of the same order as an easily-obtained assumption $(t_{ox})^{-1} = 3.3 \times 10^{-3}$, where $t_{ox}$ is the gate oxide thickness, which is the simple model of the electric field $V_{ds}/t_{ox} \approx \alpha V_{ds}$.[53] These results indicate that the observed luminescence can be well explained by the impact excitation mechanism from the theoretical point of view. Moreover, compared with the EL emitter with an individual SWNT in our previous study,[13] the threshold voltage of the light emission is reduced by approximately 80% (~ 20 V for an individual-SWNT EL[13] and ~ 4 V in this study). The reason is that the light emission is preferentially obtained from the SWNTs with small contact resistance in this SWNTs thin film, and the threshold voltage of the emitter with SWNTs thin films is lower than that with an individual SWNT.

The fact that the EL emission is obtained from the SWNTs can be confirmed by observing the polarization dependence of the light emission, because the polarization direction of EL emission is parallel to the axial direction of a SWNT, as was previously reported.[5,18,20] Figure 3a shows the polarization dependence of the emission spectrum. Although the total emission intensity of all the measured wavelength regions shows weak polarization dependence, strong polarization dependence can be observed at each of the peaks that correspond to different chiralities. Because each chirality SWNT in the prepared thin film is randomly oriented, the magnitudes of the polarization angles at each of the peaks do not show any particular manner. In the luminescence image obtained by a NIR camera, shown in Fig. 3b, the strong polarization characteristic can be observed at the localized small emission spot, whereas the polarization characteristic is weak at the large and bright emission spot (Fig. 3b). For the localized small emission spot, polarization dependence of the emission intensity is given by $\cos^2\theta$, where $\theta$ is the azimuth angle of a polarizer, and the intensity becomes zero at $\theta = 90°$ when the emission is from an individual SWNT. On the other hand, for the large and bright emission spot, the emission intensity does not become zero at any angle and shows weak polarization dependence, because the large and bright spot consists of many randomly oriented SWNTs.

We measured the time-resolved EL emission to demonstrate the high-speed modulation of the EL emitters and elucidate the relaxation time of the EL emission. Figure 4 shows the time-resolved measurements of the emission intensity when the rectangular inputs with the widths of 1, 5, and 10 ns, height of 2 V, and 10–90% response time of 730 ps were applied. A rapidly rising and falling rectangular EL emission response to the applied voltage was observed, and the 10–90% response time of EL emission was measured as 480 ps. This response time is less than that of the input voltage because the emission intensity increases superlinearly with the increase of the input voltage, as shown in Fig. 2c and equation (1). Figure 4c shows the ideal EL response estimated from both the applied voltage waveform in Fig. 4a and fitted voltage dependence of the EL intensity with equation (1) in Fig. 2c. The ideal time-resolved EL behavior is roughly in agreement with the experimental results. However, the ideal EL response time estimated from the input waveform is 350 ps, which is slightly faster than the experimental result, as shown in Fig. 4d. To obtain the intrinsic EL relaxation time of this emitter from the experimental result, the experimental result was fitted with the theoretical EL response curve calculated by convolving the ideal EL response curve from the voltage waveform (shown in Fig. 4c) and simple relaxation curve given by $\exp(-t/\tau)$. From this fitting, the relaxation time was obtained as $\tau = 140$ ps, as shown in Fig. 4d. As this relaxation time is almost similar to the intrinsic jitter of the photodetector used in this study, it is impossible to directly obtain the relaxation time of the device just from the relaxation curve. However, taking jitter into account, these results indicate that the trion-based EL emission shows less than 140-ps relaxation time, which is consistent with the past reports on PL-trion relaxation time.[41,54,55]

In addition, we demonstrated the pulsed light generation by applying pulsed voltage with the pulse generator. Under the pulsed voltage with the 120-ps width and 3.3-V amplitude, the pulsed EL emission with 120-ps FWHM was generated, as shown in Fig. 5a. We also demonstrated the high-speed continuous modulation of the EL emission. Under the modulated input voltage at 1 GHz, the EL emission quickly responds to the input signal at 1 GHz, as shown in Fig. 5b. Taking the relaxation time of this EL emitter (140 ps) into account, the high-speed modulation at several GHz is theoretically possible. This indicates that the SWNT-based EL emitter can be applied as the light source for optical communications on silicon chips.

**Conclusions**

We have demonstrated the high-brightness high-speed ultra-small-size EL emitters directly fabricated on silicon substrates by using semiconducting SWNTs thin films, which may lead to new optoelectronic integrated circuits on silicon chips. The SWNT-based EL emitters have advantages compared with the light emitters based on compound semiconductors because the SWNT EL emitters do not require p-n junctions, rare metals, or harmful substances. In addition, it is also expected to provide a new optical interconnect, which does not require such an optical modulator as the high-speed modulatable EL emitters can be highly integrated. Compared with black-body radiation CNT emitters, EL emitters show higher luminous efficiency. As the response speed of the EL emission is not affected by the substrate (black-body radiation is affected by the substrate), suspended SWNTs can increase the luminous efficiency. The brightness can be further increased by confining excitons or doping oxygen. In addition, both the response speed and light emission efficiency can be controlled by manipulating the defects in the SWNTs. Usually, the EL emission from SWNTs is sufficient to obtain a specific wavelength for optical communications. Moreover, the sharp peak shape of the EL emission compared with the blackbody emitters is advantageous to reduce the chromatic dispersion.

**Methods**

**Device Fabrication.** We used SWNTs mainly having (7,6) chirality ($d = 0.93 \pm 0.27$ nm CoMoCAT SWeNT SG76) purchased from E&T Co. for our devices. In addition to (7,6)-chirality SWNTs, other (8,6), (8,7), (7,5), and (10,2)-chirality SWNTs and slightly metallic CNTs were also included. First, 1 wt% of the surfactant (sodium dodecyl sulfate) was dissolved in water as a solvent for dispersing CNTs. Powdered CNTs were also dissolved and dispersed for approximately 1 h by using ultrasonic homogenizer. Then, the CNT dispersion was dropped on the Si substrate, which had a 30 × 30 μm square hole formed by using electron beam (EB) lithography with resist (ZEP520A). The following process was repeated 5 times: the substrate was kept tilted in various directions for approximately 1 min, the CNT dispersion was spin-coated, and the specimen was baked at 120 °C for 1 min using a hot plate. After that, the specimen was dipped in a dichloromethane solution for 5 min to lift-off and then rinsed with acetone. The DC devices were fabricated using the EB lithography and lift-off process on a heavily doped p-type Si substrate with a 300-nm-thick thermally oxidized $SiO_2$ layer. After forming the SWNTs thin film, the source and drain Pd electrodes with the gap of 500 nm and width of 30 μm were formed on the semiconducting SWNTs thin film (Fig. 1b). This DC device emits light by applying a bias voltage between the source and drain and can apply a back gate voltage via a substrate. The high-speed devices were fabricated by the same procedure as the DC devices, except for the electrode shapes and substrate type. A coplanar waveguide Pd/Ti electrode with the characteristic impedance of 50 Ω was formed on an undoped Si substrate. A rectangular or pulse voltage was applied to the signal line to obtain the high speed modulated EL emission in the high-speed modulation measurements.

**Optical Measurements.** The emission from SWNTs thin film EL devices was measured by a micro-photoluminescence measurement system, in which the emitted light was collected through a quartz optical window of a high-vacuum sample chamber and microscope objective lens at room temperature, as described in the previous reports.[2,4,13,24] 2D emission images and emission spectra were obtained by an InGaAs CCD camera and InGaAs linear detector with a spectrometer with approximately 0.8–1.6 μm detection wavelength range, respectively. The relative spectral response of the measurement system, including the optical path and detector, was measured with a standard light source blackbody furnace, and all the spectra were corrected accordingly. An IR polarizer was inserted between the sample and CCD camera (or spectrometer) for the emission polarization measurements. In general, optical elements centered on grating used for spectroscopy have strong polarization dependence. To obtain the true emission spectrum from the SWNTs thin film emitters, the polarization dependence of the light detection efficiency was measured, and the results were calibrated accordingly.

In the high-frequency measurements, bias signals were applied through a signal electrode of the high-frequency coplanar device using a DC voltage source, signal generator, function generator, and pulse generator for steady-state emission, time-resolved high-speed emission measurements, real-time optical communications, and pulse light generation, respectively. In the time-resolved emission measurements based on a single-photon counting method, the emitted light was guided to the Geiger-mode InGaAs Alanche photodiode (APD) with the wavelength range from 0.8 to 1.7 μm through a fiber coupler and multimode optical fiber. The time-resolved emission intensity was obtained using a time-correlated single-photon counting module, in which the time differences between the detected photon signals from the APD and synchronizing signals from the signal or pulse generator were measured. The time-resolved measurement based on a single-photon counting method is an effective method to investigate the response speed of the low-intensity light sources, such as SWNT-based emitters, as the emission response can be accurately measured independently of the light intensity. As all the optical measurements were carried out using InGaAs-based detectors (the CCD camera, linear array detector and Geiger-mode APD), the measurement results are suitably matched for the telecommunication wavelength band with fiber optics (1.26–1.63 μm).

FIGURES

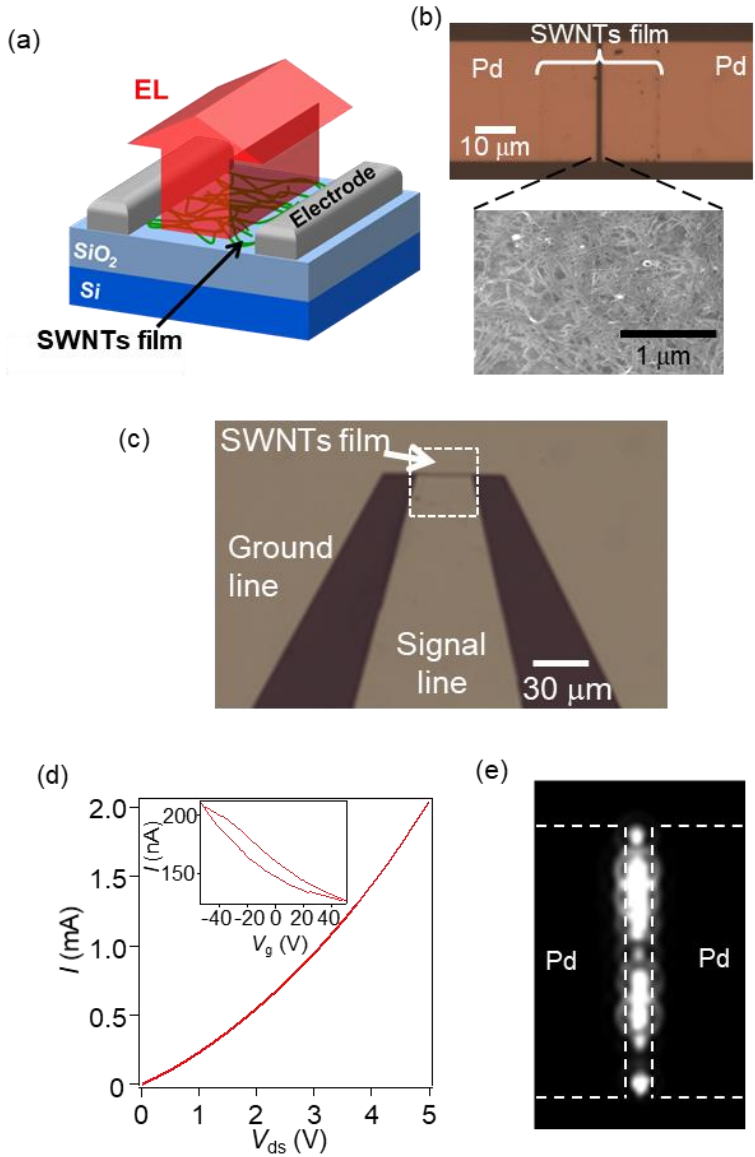

Figure 1. (a) Schematic illustration of an EL emitter. A spin-coated SWNTs thin film was formed on a SiO$_2$/Si substrate between two terminal Pd electrodes. The EL emission is obtained by applying bias voltage ($V_{ds}$). (b) Optical microscope and SEM images of the DC device. The thin film with randomly oriented SWNTs is directly formed on the substrate. (c) Optical-microscope

image of the high-speed device with a coplanar waveguide. White dashed line indicates the square-shaped SWNTs thin film. (d) $V_{ds}$ dependence of the current $I$ at a gate voltage $V_g = 0$ V for the DC device. The current shows slight superlinear behavior because of the current increase by the reduction of the Shottky barrier at high voltage. The inset shows the $V_g$ dependence of the current $I$ at constant $V_{ds} = 10$ mV. It has p-type characteristic, in which the current increases in the negative $V_g$ region. (e) NIR camera image of the EL emission at $V_{ds} = 5$ V for the DC device shown in (b).

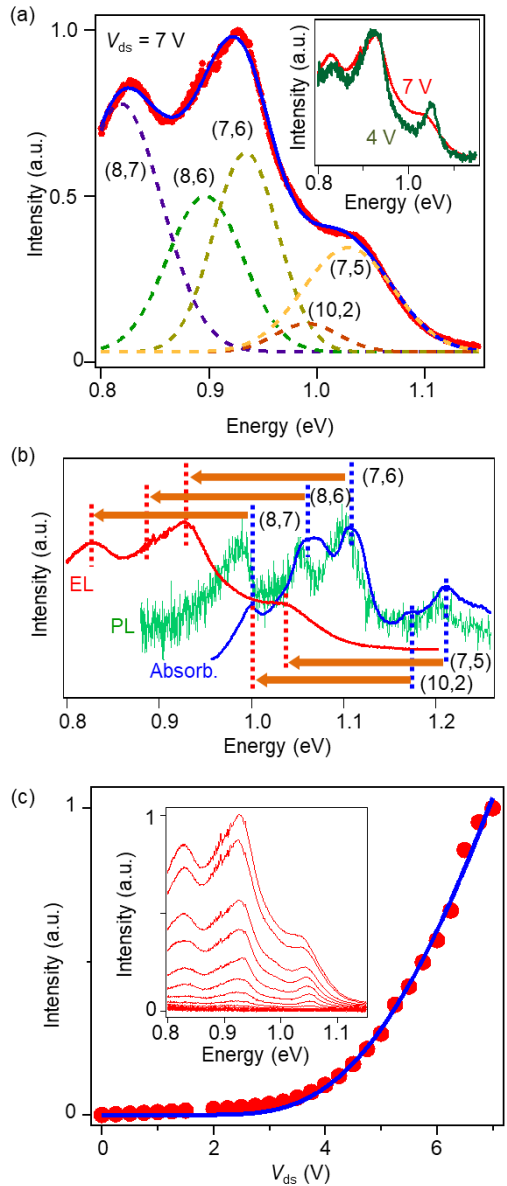

Figure 2. (a) EL emission spectrum at $V_g = 0$ V under $V_{ds}$. The blue curve indicates the fitted emission spectrum by the superposition of the five Gaussian functions, which correspond to each of the peaks of SWNTs emission with different chiralities (indicated by the dashed curves). The inset shows that the FWHM of the emission peak slightly increases with increasing $V_{ds}$ from 4 V (green curve) to 7 V (red curve). (b) Absorption spectrum (blue curve), PL spectrum excited by a 632.8 nm He-Ne laser (green curve), and EL spectrum under $V_{ds}$ = 7 V (red curve) of the SWNTs thin film. The peaks of the EL emission are 0.2-eV red-shifted from the peaks of the absorption and PL emission, which suggests the emission from trions. (c) $V_{ds}$ dependence of the total emission intensity from the EL emitter. The emission intensity increases exponentially with increasing $V_{ds}$, which is fitted with equation (1) in the main text (blue curve). The inset shows the emission spectrum at $V_{ds}$ = 0–7 V in steps of 0.25 V.

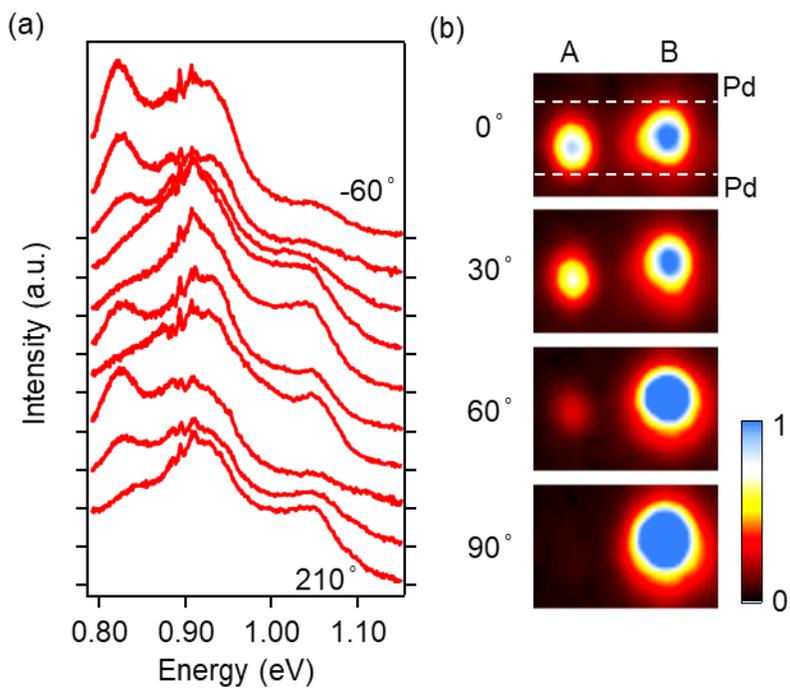

Figure 3. (a) Polarization dependence of the emission spectra. (b) Polarization dependence of the luminescence image obtained by a NIR camera.

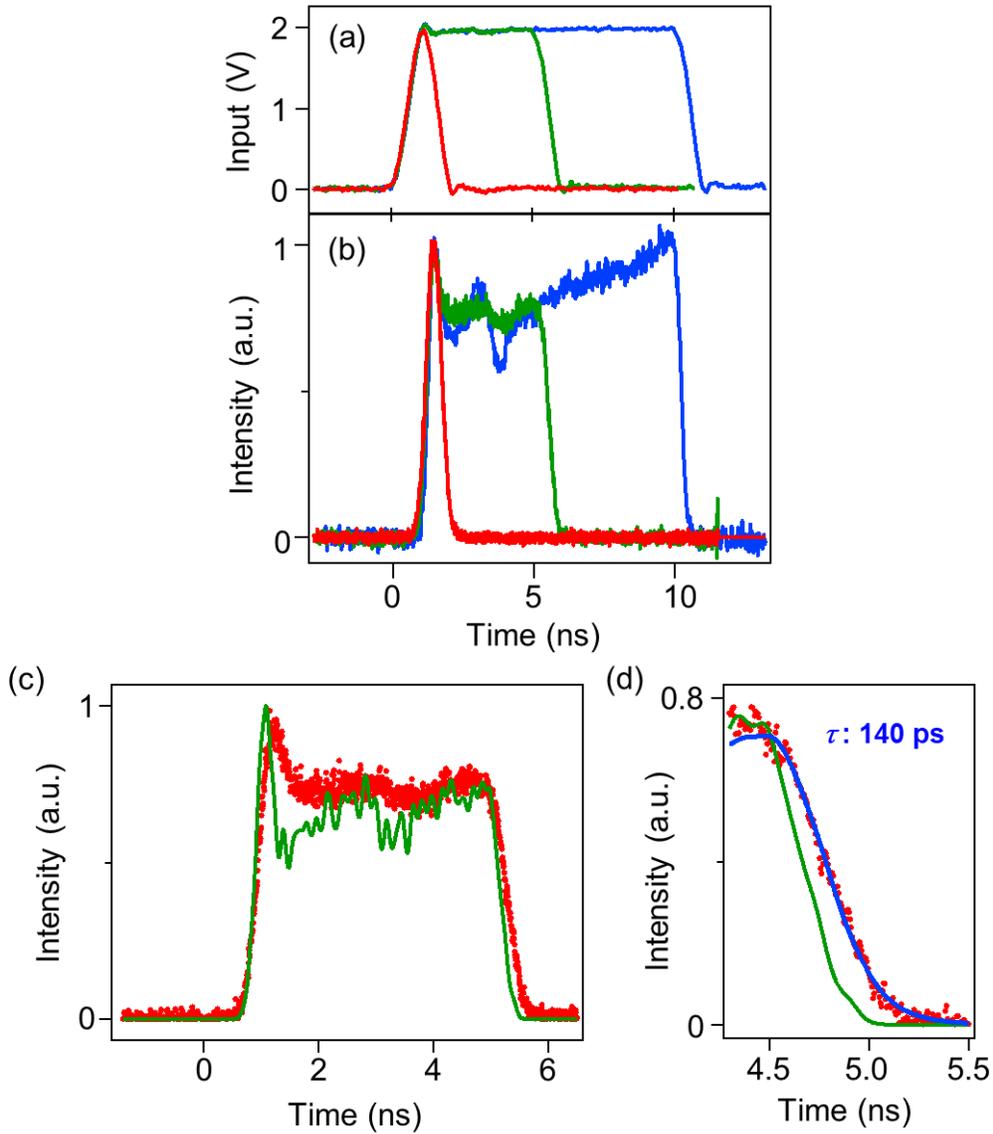

Figure 4. (a) Waveform of the applied rectangular input (1, 5, and 10-ns width and 2-V height) measured by a digital oscilloscope. (b) Time-resolved emission intensities measured by the single-photon counting method under a rectangular bias voltage shown in (a). (c) Ideal EL response estimated from both the applied voltage wave form and fitted voltage dependence of the EL intensity. Red dots represent the measured time-resolved emission intensity under a rectangular bias voltage with 5-ns width and 2-V height; green curve represents the ideal time-resolved

emission theoretically calculated using both the applied voltage waveform shown in (a) and voltage dependence of EL intensity fitted with equation (1) [shown in Fig. 2 (c)]. (d) Expanded view of (c), showing the relaxation curve of the EL emitter. The experimental result is fitted with the theoretical EL response curve (blue line) calculated by convolving the ideal EL response curve shown in (c).

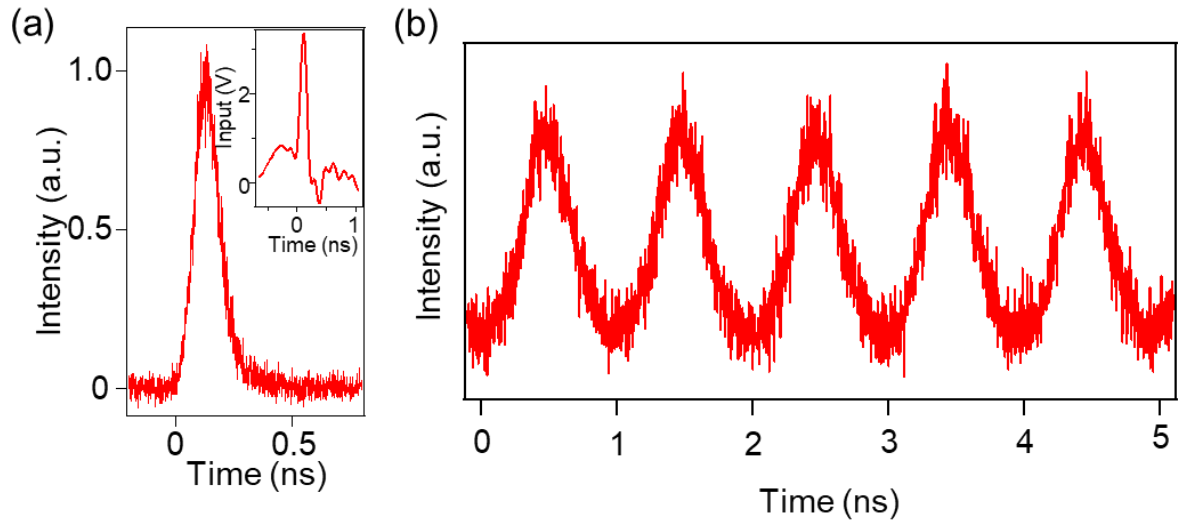

Figure 5. (a) Very-short-pulsed light emission with a width of 120 ps under a pulsed input voltage with a width of 120 ps and amplitude of 3.3 V. (b) High-speed emission modulation from the EL emitter under a continuous input (3–5 V in height) of 1 GHz obtained by time-resolved emission measurements.

AUTHOR INFORMATION

**Corresponding Author**

*Address correspondence to maki@appi.keio.ac.jp

**Author Contributions**

‡These authors contributed equally.

ACKNOWLEDGMENT

This work was partially supported by PRESTO (Grant Number JPMJPR152B) from JST, a project of Kanagawa Institute of Industrial Science and Technology (KISTEC), KAKENHI (Grant Number 16H04355, 23686055 and 18K19025) and Core-to-Core program from JSPS, Spintronics Research Network of Japan, and NIMS Nanofabrication Platform in Nanotechnology Platform Project by MEXT.